\documentclass[12pt]{article}
\usepackage{amsmath,amsfonts,graphicx,hyperref}

\begin{document}

\numberwithin{equation}{section}

\begin{titlepage}
\hbox to \hsize{\hspace*{0 cm}\hbox{\tt }\hss
   \hbox{\small{\tt }}}

\vspace{1 cm}

\centerline{\bf \Large Component twist method for higher twists in D1D5 CFT}

\vspace{1 cm}

\vspace{1 cm}
 \centerline{\large Zaq Carson $^\dagger$\footnote{zcarson@physics.utoronto.ca}\,,
Ian T. Jardine$^\dagger$\footnote{jardinei@physics.utoronto.ca}\,, and
Amanda W. Peet$^{\dagger\S}$\footnote{awpeet@physics.utoronto.ca}}

\vspace{0.5cm}

\centerline{\it ${}^\dagger$Department of Physics, University of Toronto, Toronto, ON M5S 1A7, Canada}
\centerline{\it ${}^\S$Department of Mathematics, University of Toronto, Toronto, ON M5S 2E4, Canada}

\vspace{0.3 cm}

\begin{abstract}

The deformation operator of the D1D5 orbifold CFT, a twist 2 operator, drives the CFT towards the black hole dual and its physics is key to understanding thermalization in the D1D5 system. To further study this deformation, we extend previous work on the effect of twist 2 operators to a method that works for higher orders, in the continuum limit. Our component twist method works by building higher twist operators out of twist 2 operators together with knowledge of Bogoliubov transformations. Consequently, this method sidesteps limitations in Lunin-Mathur technology by avoiding lifts to the covering space. We verify the method by reproducing results obtainable with Lunin-Mathur technology. Going further, our method upholds a previously conjectured scaling law in the continuum limit that applies to any generic configuration of twists. We illustrate this with computations for a new configuration of two twist 2 operators that twists three copies together. 

\end{abstract}

\end{titlepage}
\section{Introduction}\label{SectionIntro}

Holography plays an important role in exploring key questions in quantum gravity such as the black hole information problem. The first concrete realization of holography was the AdS/CFT correspondence of string theory, invented by taking near-horizon limits of large numbers of certain types of branes \cite{Maldacena:1997re}. Since then, other holographic dualities have been engineered using systems of branes, including classes with less supersymmetry, but this may not be the most efficient way to discover holographic dualities in general. Another approach is to start by inspecting CFTs and asking what conditions a CFT should satisfy in order for it to have a well defined AdS gravity dual. These include a large central charge and a sparse spectrum of light operators  \cite{Heemskerk:2009pn}. With these assumptions, useful information can be extracted in the context of quantum gravity to help address questions about black hole physics, for example universalities arising via $1/c$ and $h/c$ expansions \cite{Hartman:2014oaa,Keller:2014xba,Benjamin:2015hsa,Anous:2016kss,Fitzpatrick:2016ive}.

Finding CFTs that satisfy holographic restrictions is not easy: they appear to be rare. One fruitful line of investigation is to focus on symmetric orbifold CFTs, which are more likely to be holographic \cite{Haehl:2014yla,Belin:2014fna,Belin:2015hwa,Belin:2016yll}. They possess the required sparse light spectrum, while also giving the desired growth at large energies consistent with existence of black holes in the dual theory. Accordingly, investigating the general structure of  orbifold CFTs and the states in them is likely to be important to understanding holography more generally.

The CFT of the D1D5 brane system is a well known prototype holographic CFT. At one point in the moduli space it is described by a free $(T^4)^N/S_N$ orbifold. Of course, the dual at this point is strongly coupled and we need to deform towards the supergravity description. The deformation operator that accomplishes this is the superdescendent of the twist 2 operator. This twisted operator introduces interactions between the copies, which is dual to thermalization in the CFT. So understanding the action of the twist operator is important in investigating how the black hole information problem might be resolved.

One physical property known for 1+1 dimensional free orbifold theories is that acting on the vacuum with twist operators will produce a squeezed state. This behavior holds even when one applies an arbitrary number of twists \cite{Henning:1995sm,Avery:2010er}. We can then try to find the matrices describing these squeezed states. An existing technique for computing quantities with twisted operators in free symmetric orbifold CFTs, the Lunin-Mathur method, involves mapping the orbifold CFT into a covering space where all fields are single-valued \cite{Lunin:2000yv,Lunin:2001pw}. These maps must later be inverted to return to physical spacetime. Eventually the maps become quintic and higher-order polynomials once sufficiently many twists are applied. Such polynomials cannot be inverted, which severely limits the utility of these techniques for higher numbers of twist operators. Going to the continuum limit, where the radius of the circle on which D1-branes are wrapped becomes large, enables one to sidestep this problem.

The effect of twist 2 operators in the D1D5 CFT has been studied before in the context of deformations towards a black hole spacetime description \cite{David:1999ec,Gava:2002xb,Avery:2010er,Avery:2010hs,Burrington:2012yq,Carson:2014yxa,Carson:2014xwa,Carson:2014ena,Burrington:2014yia,Gaberdiel:2015uca,Carson:2015ohj,Carson:2016cjj,Carson:2016uwf,Burrington:2017jhh}. We continue the investigations begun there to study general twisted states, specifically in the continuum limit. Our method, using twist 2 operators to build up general twist configurations, reproduces previous results obtainable using Lunin-Mathur technology in the continuum limit. We also compute a new second-order twist 2 configuration and find that the coefficients defining the squeezed state scale as conjectured in \cite{Carson:2016cjj}. Our methodology gives further evidence that these general scalings should hold in the continuum limit for arbitrary configurations of twist operators.

The remainder of this paper is organized as follows. In Section \ref{SectionMethod}, we present the method in detail for twist operators in orbifold CFTs. In Section \ref{SectionD1D5} we discuss briefly the particular orbifold CFT arising from the D1D5 black hole construction.  We also discuss here the specific supersymmetry relations that relate the bosonic and fermionic sectors of this theory. In the subsequent two sections, we apply our method to two different twist configurations in the D1D5 CFT. In Section \ref{SectionOneLoop} we tackle a previously-studied configuration and find good agreement with known results, while in Section \ref{SectionNewResult} we tackle a new twist configuration. Section \ref{SectionDiscussion} concludes with a few closing remarks. Our work builds heavily on \cite{Carson:2014xwa}, which presented the basics of this method entirely in the context of the D1D5 CFT.

\section{Overview of the Method}\label{SectionMethod}
Given a free orbifold CFT in $1+1$ dimensions, consider any arbitrary combination of individual twist operators $\sigma_n$, which together we will call $\hat{\sigma}$. This operator takes a free theory living on one set of windings to a free theory on another set of windings, with the total winding number remaining constant. For the moment, consider a scalar field $\phi$ exists on all windings. The allowed excitation modes for this field will differ between the pre-and post-twist configurations. The result is 
\begin{eqnarray}
\hat{\sigma}|0\rangle &=& C\,\text{exp}\left[\sum_{m,n>0} \sum_{(j)(j')}\gamma^{(j)(j')}_{mn}a^{(j)\dagger}_m a^{(j')\dagger}_n\right]|0'\rangle ~\equiv~ |\chi(\hat{\sigma})\rangle\nonumber \\
\hat{\sigma}a^{(i)\dagger}_m|0\rangle &=& \sum_{n>0}\sum_{(j)} f^{(i)(j)}_{mn} a^{(j)\dagger}_n |\chi(\hat{\sigma})\rangle. \label{GeneralTwistForm}
\end{eqnarray}
The new parenthetical indices refer to the various CFT copies of the theory. For clarity, we use the index $(i)$ exclusively for CFT copies before the twist operators and the index $(j)$ exclusively for copies after the twist operators. The index $(k)$ will be used when speaking of an arbitrary copy without specifying one side side of the twist operators (including copy structures between the twist operators). Primes are used for additional copies of the same type. For calculational convenience, the mode indices $m$ and $n$ indicate the energy of each excitation. They can be fraction-valued for multiwound copies. The excitation level of each excitation is its mode index times the winding number of the CFT copy on which the excitation lives.

A method for handling the bosonic sector of these twist operators was explored in \cite{Carson:2014xwa} in the context of the D1D5 CFT, though the need to invert infinite matrices inhibited any useful generalization. We present a straightforward workaround for obtaining the required inverse matrices. For the remainder of this paper, we will work in the continuum limit of our orbifold CFT.

\subsection{Bogoliubov approach for arbitrary orbifold CFTs}
Consider a particular component CFT copy $(i)$ with winding number $N_{(i)}$ before the twist insertions. On this copy, the field $\phi$ may be expanded as:
\begin{eqnarray}
\phi(x) &=& \sum_{m N_{(i)}\in \mathbb{Z}^+} \left(h^{(i)}(x) a^{(i)}_m + h^{(i)*}_m(x) a^{(i)\dagger}_m \right),\label{PreTwistExpansion}
\end{eqnarray}
Now consider a particular copy $(j)$ with winding number $N_j$ after the twist insertions. Here $\phi$ may be expanded as:
\begin{eqnarray}
\phi(x) &=& \sum_{m N_{(j)}\in \mathbb{Z}^+} \left(h^{(j)}(x) a^{(j)}_m + h^{(j)*}_m(x) a^{(j)\dagger}_m \right).\label{PostTwistExpansion}
\end{eqnarray}
As described in \cite{Carson:2014xwa}, these two expansions share a linear relationship:
\begin{eqnarray}
a^{(i)}_m &=& \sum_{nN_{(j)} \in Z^+}\left(\alpha^{(i)(j)}_{mn} a^{(j)}_n + \beta^{(i)(j)}_{mn}a^{(j)\dagger}_n\right). \label{LinearRelationship}
\end{eqnarray}
The two matrices appearing here are also related to the coefficients of (\ref{GeneralTwistForm}).
\begin{eqnarray}
f^{(i)(j)}_{mn} &=& \left(\alpha^{(i)(j)}\right)^{-1}_{nm}\nonumber \\
\gamma^{(j)(j')}_{mn} &=& \sum_{kN_{(i)}\in Z^+}\sum_{(i)}\left(\alpha^{(i)(j)}\right)^{-1}_{mk}\beta^{(i)(j')}_{kn}.
\end{eqnarray}
It will be convenient to occasionally write these relations in their matrix form.
\begin{eqnarray}
f = \left(\alpha^{-1}\right)^T, \qquad \gamma = \alpha^{-1}\beta = f^T\beta.
\end{eqnarray}

The Bogoliubov matrices, $\alpha$ and $\beta$, are straightforward to calculate. They can be expressed in terms of the positive frequency solutions to the wave equation used in the expansions (\ref{PreTwistExpansion}) and (\ref{PostTwistExpansion}). The functions $h$ form a complete orthonormal basis with respect to the inner product:
\begin{equation}
(h,g) \equiv -i \int_{\Sigma}d\Sigma^{\mu}(f\partial_{\mu}g^* - g^*\partial_{\mu}f),
\end{equation}
where $\Sigma$ is the Cauchy hypersurface where the domains of $f$ and $g$ overlap. Applying this orthonormality to (\ref{PreTwistExpansion}) and (\ref{PostTwistExpansion}) and comparing to (\ref{LinearRelationship}), one finds:
\begin{eqnarray}
\alpha^{(i)(j)}_{mn} &=& \left(h^{(i)}_m,h^{(j)}_n\right)\nonumber\\
\beta^{(i)(j)}_{mn} &=& \left(\left(h^{(i)}_m\right)^*\!\!\!,h^{(j)}_{n}\right).
\end{eqnarray}

\subsection{Obtaining $\alpha^{-1}$}
Once one has calculated the Bogoliubov matrices, it is necessary to then invert $\alpha$. This task is far from trivial. The Bogoliubov matrices are infinite matrices, so there is no mathematical guarantee that an inverse even exists. Fortunately $\alpha^{-1}$ gives the unique physical transition matrix $f$. Rather than inverting $\alpha$ directly, we present a direct method of computing $f$ for arbitrary twist configurations based on the behaviour of the single component twists. In principal one then need calculate only the $\beta$ matrix to proceed.

Let us consider the twist configuration $\hat{\sigma}$ again. While we allowed this configuration to contain any combination of twists $\sigma_n$, each such component can itself be written in terms of two-twist operators $\sigma_2$. We thus decompose $\hat{\sigma}$ in terms of these two-twist operators.
\begin{eqnarray}
\hat{\sigma} &=& \prod_{q=1}^Q\sigma_{2}^{\{k_q\}}(x_q),\label{SigmaDecomposition}
\end{eqnarray}
where each $\{k_q\}$ is an ordered pair of component CFTs. Since two-twist operators do not commute, it is important to specify that we will choose indices such that the twists act in order of increasing $q$.

We now act on a state containing only a single excitation of $\phi$. The left hand side is given by (\ref{GeneralTwistForm}). On the right, we will pass each twist through one at a time.
\begin{eqnarray}
\hat{\sigma}a^{(i)\dagger}_m|0\rangle &=& \sum_{n,(j)}f^{(i)(j)}_{mn}a^{(j)\dagger}_n|\chi(\hat{\sigma})\rangle \nonumber \\
&=& \prod_{q=1}^Q\sigma_{2}^{\{k_q\}}(x_q)a^{(i)\dagger}_m|0\rangle \nonumber \\
&=& \left(\prod_{q=2}^Q\sigma_{2}^{\{k_q\}}(x_q)\right)\sum_{n_1,(k_1)}f^{(i)(k_1)}_{m,n_1}a^{(k_1)\dagger}_{n_1}|\chi_1\rangle \nonumber \\
&=& \left(\prod_{q=3}^Q\sigma_{2}^{\{k_q\}}(x_q)\right)\sum_{n_2,(k_2)}\left(\sum_{n_1,(k_1)}f^{(k_1)(k_2)}_{n_1,n_2}f^{(i)(k_1)}_{m,n_1}\right)a^{(k_2)\dagger}_{n_2}|\chi_2\rangle \nonumber \\
&=& \left(\prod_{q=4}^Q\sigma_{2}^{\{k_q\}}(x_q)\right)\sum_{n_3,(k_3)}\left(\sum_{n_2,(k_2)}f^{(k_2)(k_3)}_{n_2,n_3}\left(\sum_{n_1,(k_1)}f^{(k_1)(k_2)}_{n_1,n_2}f^{(i)(k_1)}_{m,n_1}\right)\right)a^{(k_3)\dagger}_{n_3}|\chi_3\rangle \nonumber \\ \nonumber \\
&\ldots& \nonumber \\ \nonumber \\
&=& \sum_{\{n_i\},\left\{(k_i)\right\}}\left(f^{(k_Q)(k_{Q-1})}_{n_Q,n_{Q-1}}\left(f^{(k_{Q-1})(k_{Q-2})}_{n_{Q-1},n_{Q-2}}\left(\ldots \left(f^{(k_2),(k_1)}_{n_2,n_1}f^{(k_1),(i)}_{n_1,m}\right)\right)\right)\right)a^{(k_Q)\dagger}_{n_Q}|\chi(\hat{\sigma}).\label{OneAtATime}
\end{eqnarray}
Here each $|\chi_i\rangle$ is the state obtained by acting with the first $i$ twists on the vacuum. The parentheses in the last line are used to present the correct order for the infinite sums. The fact that this order matters is equivalent to the non-associativity of infinite matrix multiplication. We have however played a little loose with our copy notation. We have used $k_i$ to indicate both a pair of copies defining a twist operator (when in brackets) and an intermediate CFT copy index (when in parentheses). 

By construction, the sum over $(k_Q)$ in the last line of (\ref{OneAtATime}) is the same as the sum over $(j)$ the first line. Similarly, the sum over $n_Q$ in the last line is the same as the sum over $n$ in the first line. We thus find:
\begin{eqnarray}
f^{(i)(j)}_{mn} &=& \sum_{\{n_i\},\left\{(k_i)\right\}}\left(f^{(j)(k_{N-1})}_{n,n_{N-1}}\left(f^{(k_{N-1})(k_{N-2})}_{n_{N-1},n_{N-2}}\left(\ldots \left(f^{(k_2),(k_1)}_{n_2,n_1}f^{(k_1),(i)}_{n_1,m}\right)\right)\right)\right).\label{fRelationSum}
\end{eqnarray}
In matrix form, this gives
\begin{eqnarray}
f\left(\hat{\sigma}\right) &=& \left(f_Q\left(f_{Q-1}\left(\ldots\left(f_2 f_1\right)\right)\right)\right),\label{fRelationMatrix}
\end{eqnarray}
where each $f_i$ is the transition matrix for the $i^{\text{th}}$ two-twist in the decomposition (\ref{SigmaDecomposition}). 

Now that we have an expression for the transition matrix, and by extension $\alpha^{-1}$, for arbitrary twist configurations, the generalization of \cite{Carson:2014xwa} is straightforward. Calculate the transition matrix and Bogoliubov matrices as above. The matrix $\alpha$ can be used to check the transition matrix calculation by verifying $\alpha f^T = f^T\alpha = \mathbb{I}$. One then finds the matrix characterizing the sqeezed state via $\gamma = f^T \beta$. To show this method explicitly, we turn to the D1D5 CFT.

\section{Application to the D1D5 CFT}\label{SectionD1D5}
The D1D5 CFT has all of the ingredients needed to apply our method. For a comprehensive review, see \cite{David:2002wn}. In fact, supersymmetry relations in the theory allow us to access most fermion modes from the bosonic sector in the continuum limit as well. In the next two sections we will look at two specific twist configurations in this theory, one with a known exact solution and one with no known exact solution. Here we introduce our notation and review the relations that allow us to probe the fermionic sector.

The D1D5 system is constructed in Type IIB string theory compactified as $M_{4,1} \times S^1 \times T^4$. The $N_5$ D5 branes wrap the full compactification, while the $N_1$ D1 branes wrap $S^1$. We take $S^1$ to be much larger than $T^4$ so that in the low energy limit only modes along $S^1$ are excited. This gives a CFT on $S^1$ with $(4,4)$ supersymmetry. We begin at the orbifold point, where the CFT is a free $1+1$ dimensional sigma model with target space $(T^4)^{N_1N_5)}/S_{N_1N_5}$. The twist operator is a part of the blow-up mode that allows us to perturbatively deform away from this orbifold point.

There are two $SO(4)$ symmetries in this theory, an exact symmetry from rotations in the four noncompact spatial directions, labelled $SO(4)_E$, and an approximate symmetry from rotations in $T^4$, labeled $SO(4)_I$. We express each of these symmetries as $SU(2) \times SU(2)$, giving four different $SU(2)$ symmetry groups. We use a different index type for each $SU(2)$ charge.
\begin{eqnarray}
SO(4)_E \to SU(2)_L \times SU(2)_R, \qquad \alpha, \dot{\alpha}\nonumber \\
SO(4)_I \to SU(2)_1 \times SU(2)_2, \qquad A, \dot{A}.
\end{eqnarray}
The bosonic and fermionic fields, along with their excitation modes, then carry the following  $SU(2)$ structure (in addition to their CFT copy index):
\begin{eqnarray}
X^{(k)}_{A\dot A} &\to& a^{(k)\dagger}_{A\dot A,n}, \, \bar{a}^{(k)\dagger}_{A\dot A,-n} \nonumber\\
\psi^{(k),\alpha A} &\to& d^{(k)\dagger,\alpha A}_n \nonumber\\
\bar{\psi}^{(k),\dot{\alpha}\dot{A}} &\to& \bar{d}^{(k)\dagger,\dot{\alpha}\dot{A}}_n.
\end{eqnarray}
Note that for bosons, the chiral primary field is actually $\partial X$.  The left-moving (holomorphic) and right-moving (anti-holomorphic) sectors largely decouple. In most cases, including the properties of bare twists, it is sufficient to work only in the holomorphic sector. The anti-holomorphic sector behaves analogously. We will therefore work predominately in the holomorphic sector for the remainder of this paper.

The theory's twist operator carries charge under the $SO(4)_E$ group, which we write as $SU(2)_L$ and $SU(2)_R$ charges. This has no effect on the bosons so we will often drop such indices from our notation. The theory also has a supercharge operator.
\begin{equation}
G^{\alpha}_{\dot{A}} = \psi^{\alpha A}\partial X_{A\dot A}.
\end{equation}
The blow-up mode that deforms the CFT away from its orbifold is an $SU(2)_L$ singlet combination of the supercharge and the twist ($SU(2)_R$ singlet in the anti-holomorphic sector). It was shown in \cite{Avery:2010er} that the two singlet charge combinations are proportional. Because of this, it is traditional to work solely with twists of positive charge. We therefore do not consider negative-charge twists in this paper.

Even restricting to only positive-charge twists, one might still expect a proliferation of $SU(2)$ indices on all of our characteristic coefficients. This is not the case. Aside from the fermion zero modes, there is only a single linearly independent $\gamma$ and $f$ matrix for each copy combination. In practice, however, it is convenient to write different matrices for bosons and fermions and to also add an $SU(2)_L$ index to the fermion transition matrix. One then has the particular $SU(2)$ structures:
\begin{eqnarray}
|\chi(\hat{\sigma})\rangle &=& C\,\text{exp}\left[\gamma^{B(j)(j')}_{mn}\left(-a^{\dagger(j)}_{++,m}a^{\dagger(j')}_{--,n} + a^{\dagger(i)}_{+-,m} a^{\dagger(j)}_{-+,n}\right)\right]\nonumber\\
&&\quad{}\times\text{exp}\left[\gamma^{F(j)(j')}_{mn}\left(d^{\dagger(j)++}_m d^{\dagger(j)--}_n - d^{\dagger(j)+-}_m d^{\dagger(j)-+}_n\right)\right]|0\rangle 
\end{eqnarray}
The transition matrices are independent of all $SU(2)$ charges except $SU(2)_L$. We thus add an index $\alpha$ to the fermionic piece.
\begin{equation}
f^{B(i)(j)}_{mn}, \qquad f^{F\alpha(i)(j)}_{mn}.
\end{equation}
Our method only allows us to calculate the bosonic coefficients directly. However, the nontrivial relationships between the fermionic and bosonic coefficients were shown in \cite{Carson:2015ohj}, 
 \cite{Carson:2016cjj}. We can thus access the fermionic sector (except zero modes).\footnote{Our inability to access fermion zero modes means that our method is insensitive to the particular $SU(2)$ structure of the Ramond vacuum. This insensitivity is expected in the continuum limit.}

Our method was derived for canonically normalized modes, but unfortunately for us it has become standard for work in the D1D5 CFT to make use of modes that are not canonically normalized. We thus provide a translation between the standard mode normalizations and our canonical normalizations.
\begin{eqnarray}
\alpha^{(k)}_{A\dot A,-n} &=& \sqrt{n} \, a^{(k)\dagger}_{A\dot A,n} \nonumber \\
\tilde{d}^{(k)}_{A\dot A,-n} &=& \sqrt{N_{(k)}}\,d^{(k)}_{A\dot A,-n},
\end{eqnarray}
where the standard modes are on left-hand side. Because of this, our characteristic coefficients are also different. For the sqeezed state we have
\begin{eqnarray}
\tilde{\gamma}^{B(j)(j')}_{mn} &=& \sqrt{mn}\,\gamma^{B(j)(j')}_{mn} \nonumber \\
\tilde{\gamma}^{F(j)(j')}_{mn} &=& \sqrt{N_{(j)}N_{(j')}}\,\gamma^{F(j)(j')}_{mn},
\end{eqnarray}
while for the transition matrix we have
\begin{eqnarray}
\tilde{f}^{B(i)(j)}_{mn} &=& \sqrt{\frac{n}{m}}f^{B(i)(j)}_{mn} \nonumber \\
\tilde{f}^{F\alpha(i)(j)}_{mn} &=& \sqrt{\frac{N_{(j)}}{N_{(i)}}} f^{F(i)(j)}_{mn}.
\end{eqnarray}
We also translate the relationships between the bosonic and fermionic coefficients from \cite{Carson:2015ohj}, \cite{Carson:2016cjj}.
\begin{eqnarray}
\tilde{\gamma}^{F(i)(j)}_{mn} &=& -m{\sqrt{mn}\over\sqrt{N_{(i)}N_{(j)}}}\gamma^{B(i)(j)}_{mn} \nonumber \\
\tilde{f}^{F-,(i)(j)}_{mn} &=&\sqrt{n\over m}\sqrt{N_{(i)}\over N_{(j)}} f^{B(i)(j)}_{mn} \nonumber \\
\tilde{f}^{F+,(i)(j)}_{mn} &=& \sqrt{m\over n} \sqrt{N_{(j)}\over N_{(i)}} f^{B(i)(j)}_{mn}.
\end{eqnarray}
In all of the above expressions, the traditional forms are written on the the left-hand side.

\section{Reproduction of $1+1 \to 2 \to 1+1$ Results}\label{SectionOneLoop}
We now look at the case of two singly wound copies twisted into one doubly wound copy at $w_1 = \tau_1 + i\sigma_1$ and then untwisted back into two singly wound copies at $w_2 = \tau_2 + i\sigma_2$. For convenience, we will use translation and rotation invariance to select $\tau_1 = \tau_2 = 0$ and $\sigma_2 > \sigma_1$. This means that our Cauchy hypersurface $\Sigma$ is a collection of $\sigma$ contours.

Since we have two singly-wound CFTs in both the initial and final states, we will place primes on the explicit final-state copy indices. So the initial copies are $(1)$ and $(2)$ while the final copies are $(1')$ and $(2')$.

\subsection{Calculating the Bogoliubov matrices}
\begin{figure}
 \begin{center}
  \includegraphics[width=1.0\textwidth]{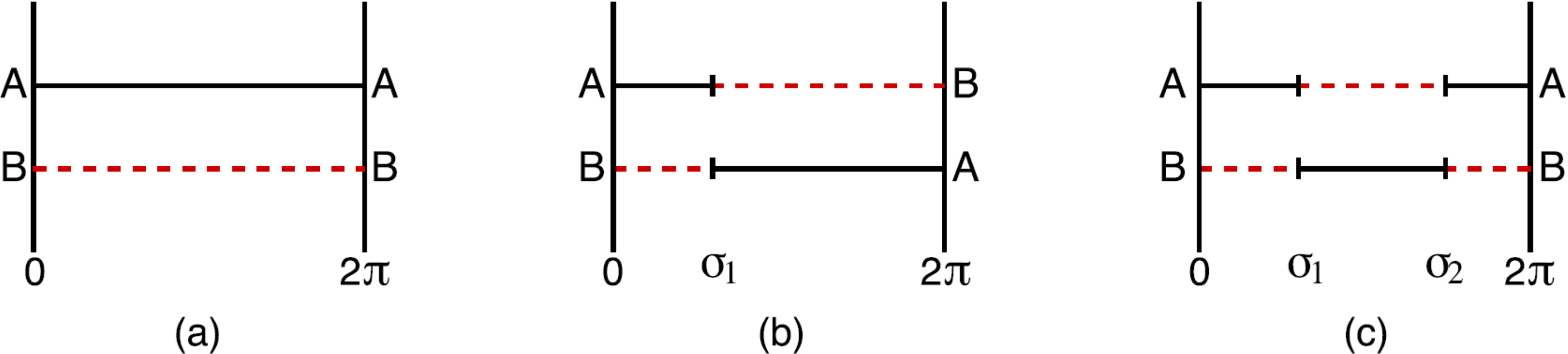}
 \end{center}
  \caption{The layout of the copies as we apply the twists. Copy (1) is the black line, copy (2) is red. Here the A,B ends are identified under the $\sigma=0\sim2\pi$ identification of the cylinder. We perform all calculations at $\tau=0$, though we separate the windings here visually for clarity. (a) Copies before twists. (b) Application of $\sigma_{(12)}$ at $\sigma_1$, twisting the copies together. (c) Application of $\sigma_{(21)}$ at $\sigma_2$ , splitting into two new copies (1') (top) and (2') (bottom)} \label{11to2to11Figure}
\end{figure}

We begin by writing the positive frequency solutions to the wave equation.
\begin{eqnarray}
h^{(k)}_m &=& {1\over \sqrt{2\pi}}{1\over \sqrt{2m}}e^{im(\sigma - \tau)}\nonumber\\
h^{(k)}_{\bar m} &=&{1\over \sqrt{2\pi}}{1\over \sqrt{2\bar{m}}}e^{-i\bar{m}(\sigma + \tau)},
\end{eqnarray}
where $m$ is an integer. The functional form is the same for all copies since each copy has the same winding number. However, different copies are valid over different regions. Importantly, copies $(1)$ and $(2)$ have no domain overlap, and similarly for copies $(1')$ and $(2')$. What we need now is the domain of overlap for each initial-final copy pair. There is in fact some ambiguity here, owing to the freedom to define our copies however we choose. We will follow the conventions of \cite{Carson:2015ohj}. The intermediate doubly-wound CFT has its interval $[0,\sigma_1]$ taken from copy $(1)$, while the second twist takes the first $2\pi$ interval of this intermediate CFT into copy $(1')$. The initial-final domains of overlap can then be seen in figure \ref{11to2to11Figure}.
\begin{eqnarray}
(1),(1') \text{ or } (2),(2') &\implies& \Sigma = [0,\sigma_1]\cup [\sigma_2,2\pi]\nonumber\\
(1),(2') \text{ or } (2),(1') &\implies& \Sigma = [\sigma_1,\sigma_2].
\end{eqnarray}
Since the expansion functions are the same for all copies, each matrix's $(1,1')$ and $(2,2')$ components are identical, as are $(1,2')$ and $(2,1')$ components. This is expected, as the physics is invariant under the combination of interchanges $1\leftrightarrow2$ and $1' \leftrightarrow 2'$.

We can now calculate our Bogoliubov matrices. Starting with $\alpha$, we have
\begin{eqnarray}
\alpha^{(1)(1')}_{mn} &=& -i \left(\int_0^{\sigma_1} + \int_{\sigma_2}^{2\pi}\right) {1\over 4\pi\sqrt{mn}}\left(e^{im(\sigma-\tau_0)}ine^{-in(\sigma-\tau_0)} \right.\nonumber\\
&&\left.\quad {}- e^{-in(\sigma-\tau)}(-im)e^{im(\sigma-\tau_0)}\right)d\sigma\nonumber\\
&=& {1\over 4\pi\sqrt{mn}}\left(\int_0^{\sigma_1} + \int_{\sigma_2}^{2\pi}\right)(m+n)e^{i(m-n)\sigma}d\sigma.
\end{eqnarray}
The result is piecewise. For $m=n$, we have
\begin{eqnarray}
\alpha^{(1)(1')}_{mm} &=& {1\over 4\pi}{1\over m}2m(\sigma_1 + 2\pi - \sigma_2)\nonumber\\
&=& 1 - {\sigma_1-\sigma_2 \over 2\pi}\nonumber\\
&=& 1 - {\Delta w \over 2\pi i}.
\end{eqnarray}
For $m \neq n$, we find
\begin{eqnarray}
\alpha^{(1)(1')}_{mn} &=& {1\over 4\pi\sqrt{mn}}{m+n\over i(m-n)}\left(e^{-i(m-n)\sigma_1} - 1 + 1 - e^{-i(m-n)\sigma_2}\right)\nonumber\\
&=& {1\over 2\pi}{m+n\over \sqrt{mn}(m-n)}e^{i(m-n){\sigma_1+\sigma_2\over 2}}{1\over 2i}\left( e^{i(m-n){\sigma_1 - \sigma_2 \over 2}} - e^{-i(m-n){\sigma_1-\sigma_2\over2}}\right)\nonumber\\
&=& {1\over 2\pi}{m+n\over \sqrt{mn}(m-n)}e^{i(m-n){\sigma_1+\sigma_2\over 2}}\sin\left((m-n){\Delta w \over 2i}\right).
\end{eqnarray}
We now use translation invariance to set the $\sigma$ midpoint to zero. This eliminates the phase.
\begin{eqnarray}
\alpha^{(1)(1')}_{mn} &=& {1\over 2\pi}{m+n\over \sqrt{mn}(m-n)}\sin\left((m-n){\Delta w \over 2i}\right).
\end{eqnarray}
As noted earlier, this is also the purely left-moving $(2,2')$ portion of the $\alpha$ matrix. The calculation for $(1,2')$ is identical up to a change in integration limits. The result is
\begin{eqnarray}
\alpha^{(1)(2')}_{mn} &=& \begin{cases}
{\Delta w \over 2\pi i} & m = n\\
-\alpha^{(1)(1')}_{mn} & m \neq n,
\end{cases}
\end{eqnarray}
which is also the purely left-moving $(2,1')$ result. As expected by their general decoupling, the purely right-moving results are identical to their purely left-moving counterparts. There are also parts of the $\alpha$ matrix with mixed holomorphicity. These are nonphysical, terms with them vanish, and serve only to illuminate the kernel of $f$. We have omitted these portions of the matrix for brevity. We thus have all the physically-relevant parts of $\alpha$.

The calculation for $\beta$ is analogous. It gives:
\begin{eqnarray}
\beta^{(1)(1)}_{mn} &=& {1\over 2\pi}{m-n\over \sqrt{mn}(m+n)}\sin\left((m+n){\Delta w\over 2i}\right)\nonumber\\
\beta^{(1)(2)}_{mn} &=& -\beta^{(1)(1)}_{mn}.
\end{eqnarray}
Conveniently, there is no need for a piecewise expression. Each component simply vanishes on the diagonal.

\subsection{Comparison to previous results}
An analytic form of the transition matrix for this twist configuration in the continuum limit was identified in \cite{Carson:2016cjj}. Shifting to canonical modes, for $0 < \Delta w \leq 2\pi i$ one has:
\begin{eqnarray}
f^{(1)(1')}_{mn} &=& f^{(2)(2')}_{mn} ~=~ \begin{cases}
\displaystyle1-{\Delta w \over 2i} & m=n\\
\displaystyle-{1\over\pi(m-n)}\sin\left((m-n){\Delta w \over 2i}\right) & m\neq n
\end{cases} \nonumber\\
f^{(1)(2')}_{mn} &=& f^{(2)(1')}_{mn} ~=~ \begin{cases}
\displaystyle{\Delta w \over 2i} & m=n\\
\displaystyle{1\over\pi(m-n)}\sin\left((m-n){\Delta w \over 2i}\right) & m\neq n
\end{cases}.
\end{eqnarray}
We can multiply this matrix by $\alpha$ to check our calculations. Using a cutoff of $\Lambda=2000$ for the sums, the result is $\mathbb{I}$ to within $0.01\%$ for mode indices of order 10 or larger.

One can also check the method of building this two-twist transition amplitude out of the behaviour of each component twist. Each component transition amplitude also scales as $1/(m-n)$, so the summand behaves as an inverse square. The convergence is fast. $\Lambda = 2000$ is sufficient for $0.2\%$ accuracy for mode indices of order 10 or larger.

Like the transition matrix, the matrix $\gamma$ has the following copy relations:
\begin{equation}
\gamma^{(1')(1')}_{mn} = \gamma^{(2')(2')}_{mn} = -\gamma^{(1')(2')}_{mn} = -\gamma^{(2')(1')}_{mn}.
\end{equation}
We therefore compute only the $(1'),(1')$ sector. This computation is more arduous. The summand scales with a power of $-3/2$ so convergence is slower. Furthermore, larger mode indices are required to maintain the validity of the continuum limit approximations. This is in keeping with \cite{Carson:2016cjj}, wherein a general from of the falloff of $\gamma$ with rising mode indices was not valid for low-lying modes.

\begin{figure}
 \begin{center}
  \includegraphics[width=0.9\textwidth]{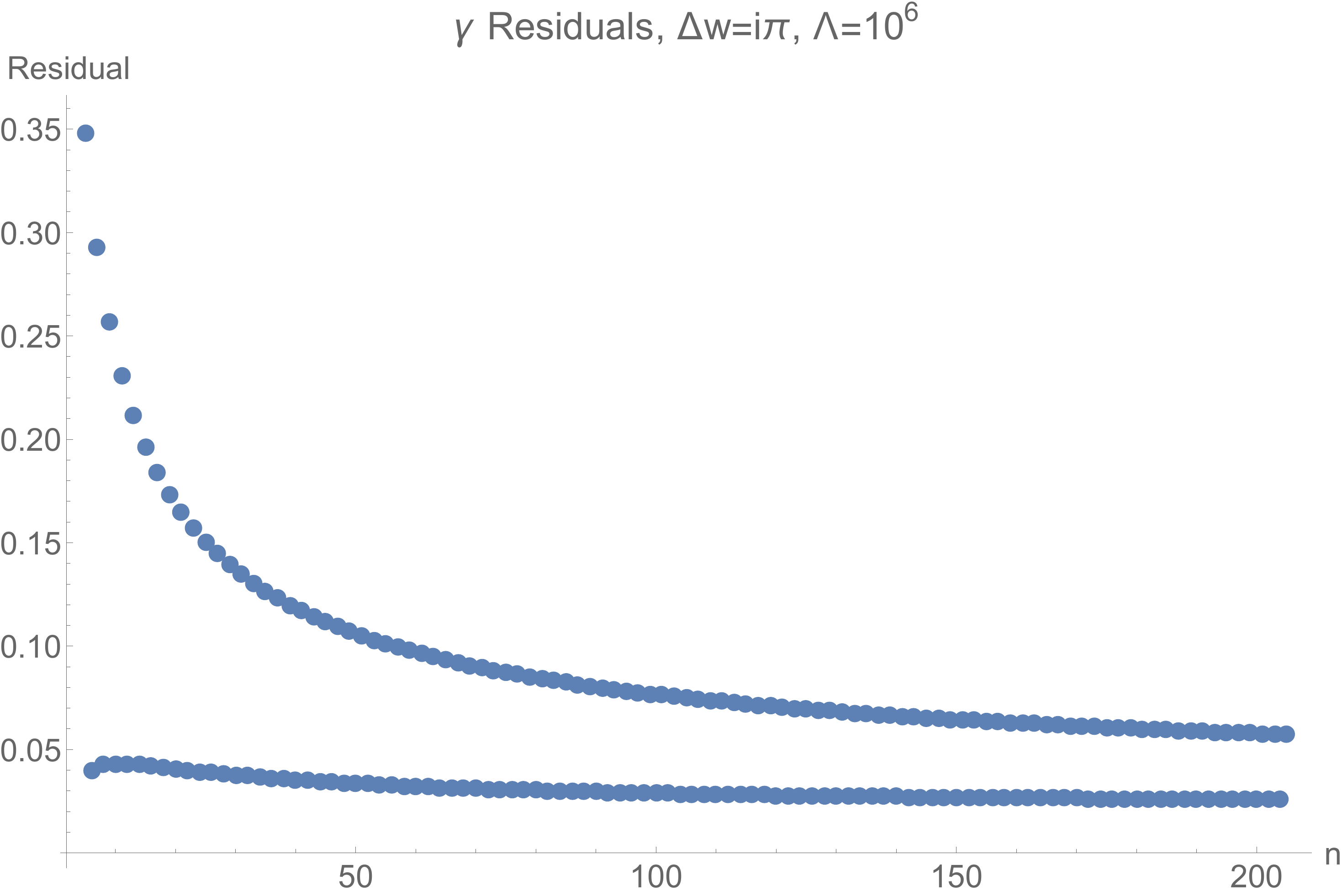}
 \end{center}
  \caption{Residual vs $n$ for $\gamma^{(1')(1')}_{n,n}$ at $\Delta w = \pi i$.} \label{GammaComparisonPlot}
\end{figure}

\begin{figure}
 \begin{center}
  \includegraphics[width=0.8\textwidth]{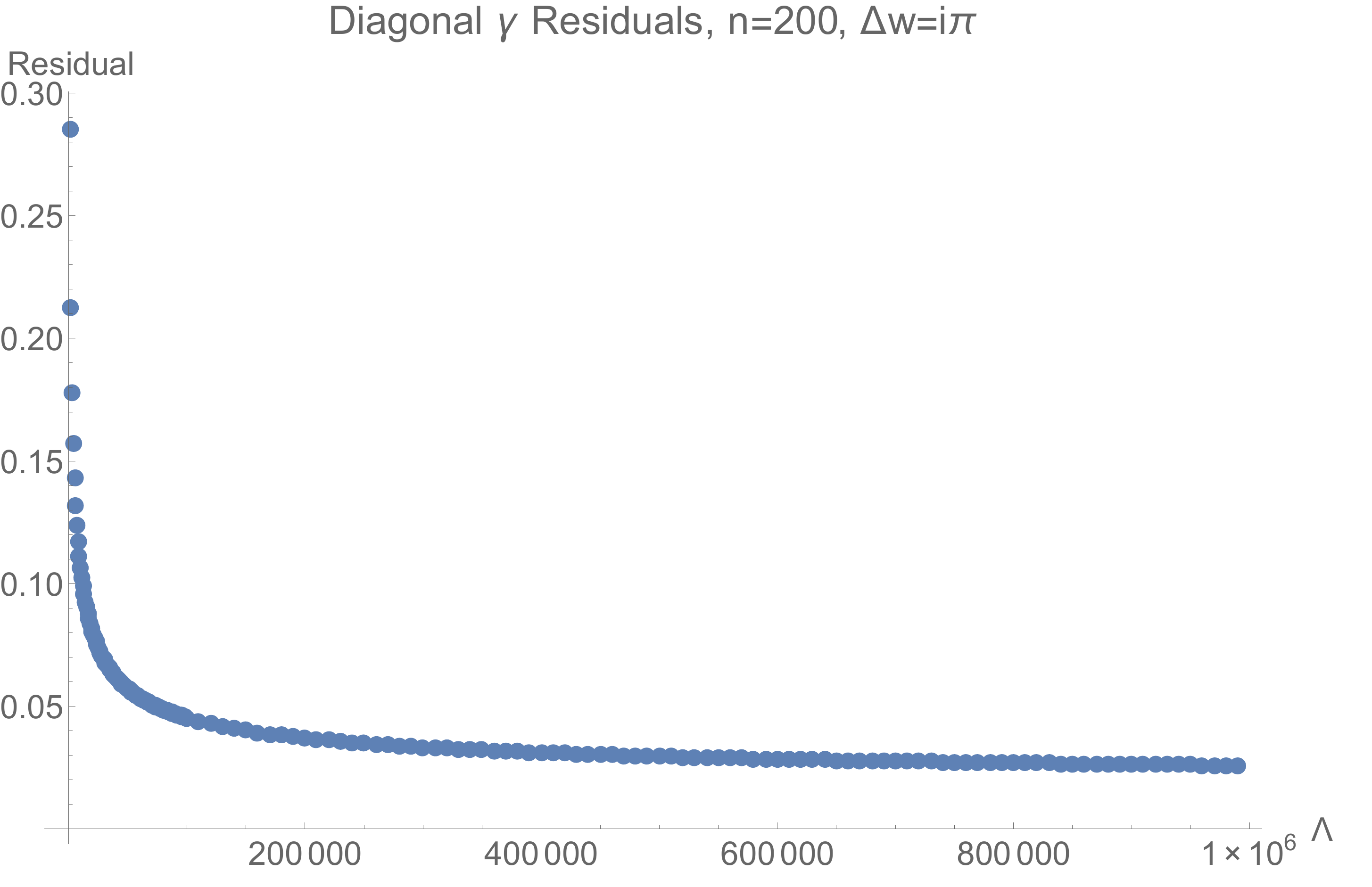}
 \end{center}
  \caption{Residual vs cutoff for $\gamma^{(1')(1')}_{200,200}$ at $\Delta w = \pi i$.} \label{ResidueCutoffRelation}
\end{figure}

\begin{figure}
 \begin{center}
  \includegraphics[width=0.8\textwidth]{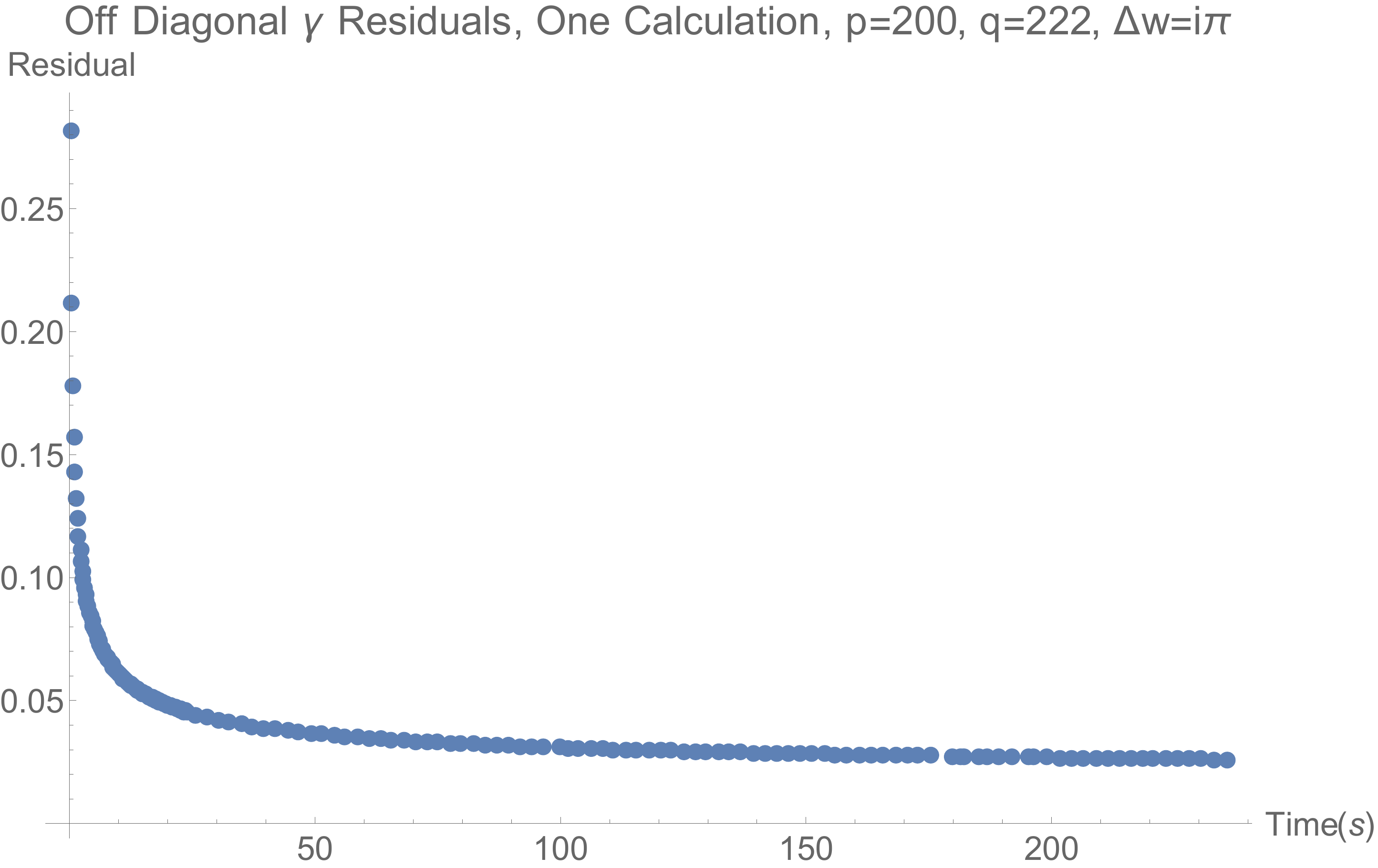}
 \end{center}
  \caption{Residual vs computation tame for $\gamma^{(1')(1')}_{200,222}$ at $\Delta w = \pi i$.} \label{ComputationScalingOffDiagonal}
\end{figure}

\begin{figure}
 \begin{center}
  \includegraphics[width=0.8\textwidth]{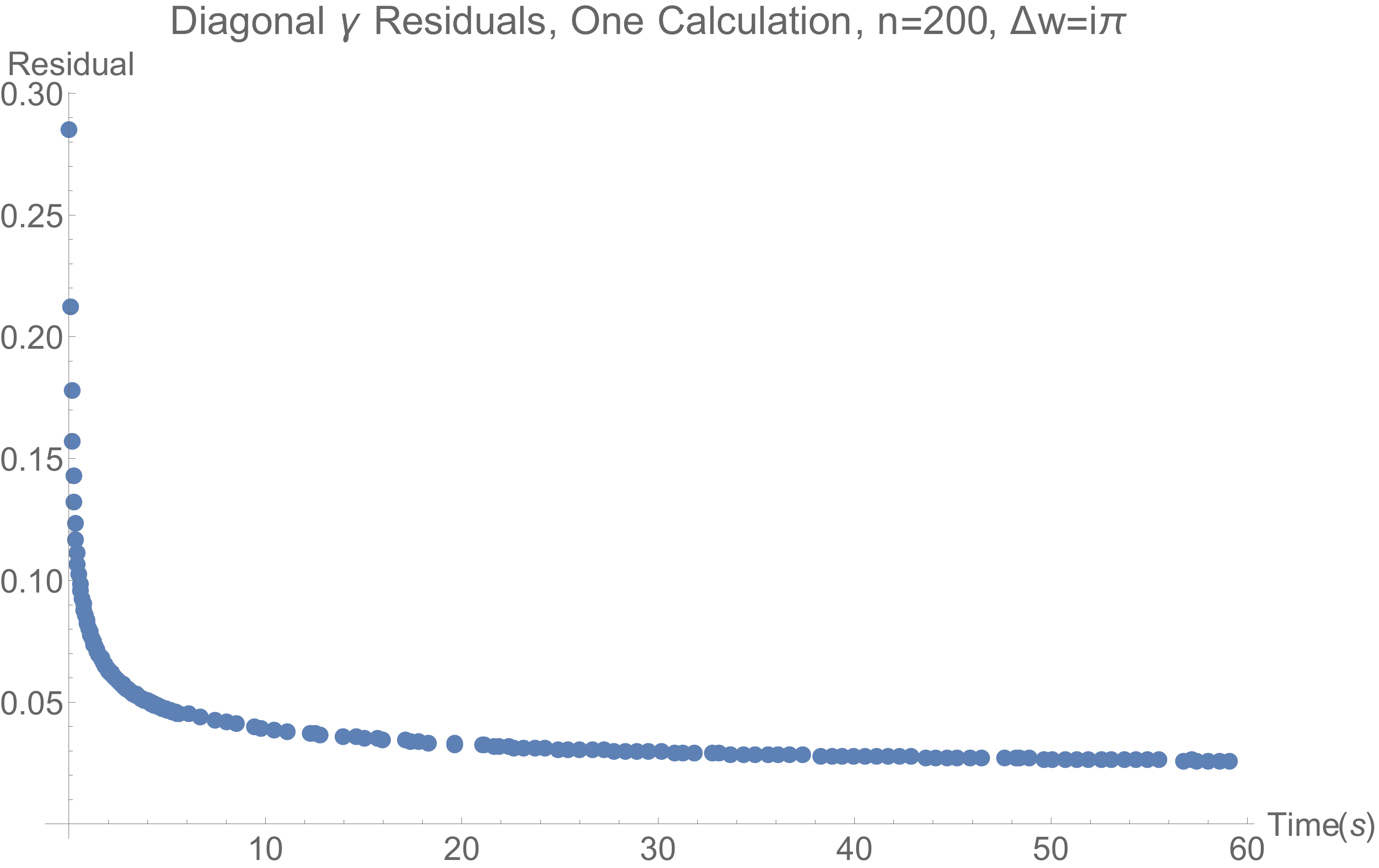}
 \end{center}
  \caption{Residual vs computation time for $\gamma^{(1')(1')}_{200,200}$ at $\Delta w = \pi i$.} \label{ComputationScalingDiagonal}
\end{figure}

Figure \ref{GammaComparisonPlot} shows the percent error of our calculation for $\Lambda = 10^6$ for various excitation levels. These high-cutoff calculations were performed on SciNet \cite{SciNetCitation}. Computations with lower cutoffs were performed on personal computers. Figure \ref{ResidueCutoffRelation} shows the cutoffs required for a desired accuracy, specifically for $\gamma^{(1')(1')}_{200,200}$. To get a feel for the accuracy/computation tradeoff, see Figures \ref{ComputationScalingOffDiagonal} and \ref{ComputationScalingDiagonal}.

\section{New Twist Configuration: $1+1+1 \to 2+1 \to 3$}\label{SectionNewResult}
We now turn to a new twist configuration which has not yet been addressed. Here we begin with three singly-wound CFT copies and through two two-twist operators join them into a single triply-wound copy. We can again use rotation and translation invariance to set $\tau_1=\tau_2=0$ and $\sigma_2 > \sigma_1$. The twist at $\sigma_1$ will join copies $(1)$ and $(2)$ while the twist at $\sigma_2$ joins copies $(2)$ and $(3)$. There is only one final copy so we drop its index.

\subsection{Calculating the Bogoliubov Matrices}
\begin{figure}
 \begin{center}
  \includegraphics[width=0.8\textwidth]{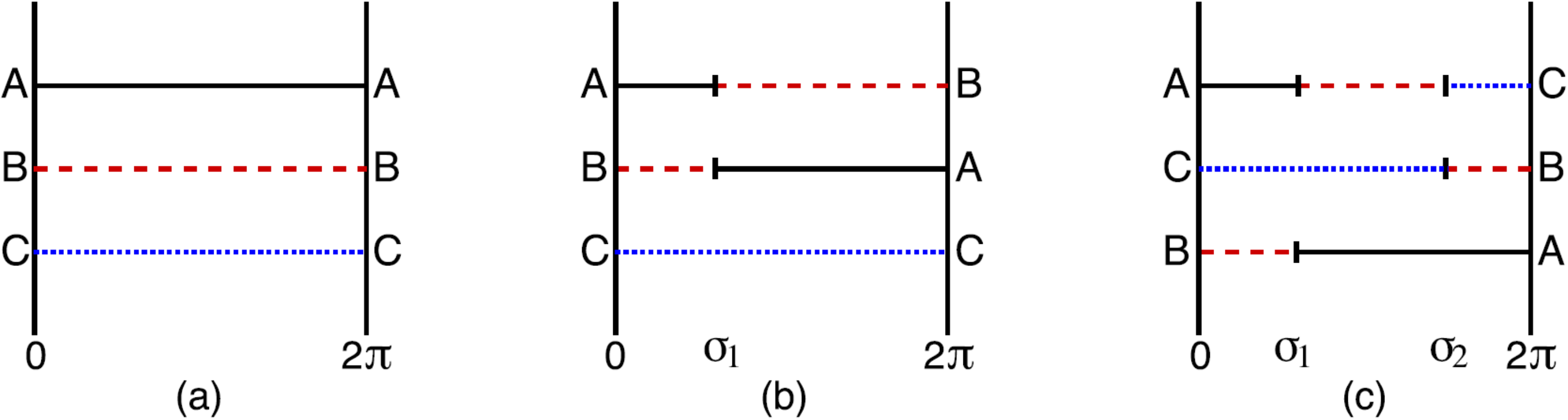}
 \end{center}
  \caption{The layout of the copies as we apply the twists. Copy (1) is the black line, copy (2) is red, copy (3) is blue. Here the A, B, C, ends are identified under $\sigma=0\sim2\pi$. We perform all calculations at $\tau=0$, though we separate the windings here visually for clarity. (a) Copies before twists. (b) Application of $\sigma_{(12)}$ at $\sigma_1$, twisting the copies together. (c) Application of $\sigma_{(23)}$ at $\sigma_2$, which twists all three together.} \label{111to21to3Figure}
\end{figure}

We begin by writing the positive frequency solutions to the wave equation. For each of the three initial copies, we have:
\begin{eqnarray}
h^{(i)}_m &=& {1\over \sqrt{2\pi}}{1\over \sqrt{2m}}e^{im(\sigma - \tau)}\nonumber\\
h^{(i)}_{\bar m} &=&{1\over \sqrt{2\pi}}{1\over \sqrt{2\bar{m}}}e^{-i\bar{m}(\sigma + \tau)},
\end{eqnarray}
with integer $m$. For the final copy, we instead have:
\begin{eqnarray}
h_{s} &=& {1\over \sqrt{6\pi}}{1\over \sqrt{2m}}e^{is(\sigma-\tau)}\nonumber\\
h_{\bar{s}} &=& {1\over \sqrt{6\pi}}{1\over \sqrt{2m}}e^{-is(\sigma+\tau)},
\end{eqnarray}
where $s$ is a multiple of $1/3$ since the copy is triply wound. We have dropped the copy index to reduce index clutter. We now identify our domains of overlap, which are diagrammed in figure \ref{111to21to3Figure}.
\begin{eqnarray}
(1) &\implies& \Sigma = [0,\sigma_1] \cup [4\pi+\sigma_1,6\pi]\nonumber\\
(2) &\implies& \Sigma = [\sigma_1,\sigma_2]\cup[2\pi+\sigma_2,4\pi+\sigma_1]\nonumber\\
(3) &\implies& \Sigma = [\sigma_2,2\pi + \sigma_2],
\end{eqnarray}
where each line is the listed copy's domain overlap with the unique final copy.

Before we proceed further, let us pause to identify an important point. Note that all of our wavefunctions are identical after shifts of $\sigma \to \sigma + 6\pi$. The domain overlap for copy $(1)$ can thus be rewritten as:
\begin{eqnarray}
(1) &\implies& \Sigma = [4\pi+\sigma_1,6\pi+\sigma_1]\nonumber\\
\end{eqnarray}
It is now clear that copies $(1)$ and $(3)$ both have contiguous domain overlaps while copy $(2)$ does not. This will manifest in significant similarity between all of copy $(1)$ and copy $(3)$ matrix components, while the copy (2) components are significantly different. This is physically expected, as copies $(1)$ and $(3)$ each only see one two-twist operator while copy $(2)$ sees both operators.

We can now proceed with the calculation of $\alpha$ and $\beta$ for each of our three initial copies. The mathematics behaves similarly to the previous case, so we present only the results. For convenience, we identify a commonly-occurring factor:
\begin{eqnarray}
\mu_s &\equiv& 1-e^{4\pi i s},
\end{eqnarray}
which vanishes for integer $s$. For $\alpha$, we again obtain a piecewise result.
\begin{eqnarray}
\alpha^{(1)}_{ms} &=& \begin{cases}
\displaystyle{1\over \sqrt{3}}\delta_{ms} & s \in \mathbb{Z}\\
\displaystyle-{i\over 4\pi \sqrt{3sm}}{s+m\over s-m}\mu_s e^{i(s-m)\sigma_1} & s \notin \mathbb{Z}
\end{cases} \nonumber \\
\alpha^{(2)}_{ms} &=& \begin{cases}
\displaystyle{1\over \sqrt{3}}\delta_{ms} & s \in \mathbb{Z}\\
\displaystyle-{i\over 4\pi \sqrt{3sm}}{s+m\over s-m}\left(\mu_{-s} e^{i(s-m)\sigma_2}-\mu_{s} e^{i(s-m)\sigma_1}\right) & s \notin \mathbb{Z}
\end{cases} \nonumber \\
\alpha^{(3)}_{ms} &=& \begin{cases}
\displaystyle{1\over \sqrt{3}}\delta_{ms} & s \in \mathbb{Z}\\
\displaystyle{i\over 4\pi \sqrt{3sm}}{s+m\over s-q}\mu_{-s} e^{i(s-m)\sigma_2} & s \notin \mathbb{Z}.
\end{cases}
\end{eqnarray}
The result for the $\beta$ matrix is:
\begin{eqnarray}
\beta^{(1)}_{ms} &=& {i\over 4\pi \sqrt{3sm}}{s-m\over s+m}\mu_{-s} e^{-i(s+m)\sigma_1} \nonumber \\
\beta^{(2)}_{ms} &=& {i\over 4\pi \sqrt{3sm}}{s+m\over s-m}\left(\mu_{s} e^{-i(s+m)\sigma_2}-\mu_{-s} e^{-i(s+m)\sigma_1}\right) \nonumber \\
\beta^{(3)}_{ms} &=& -{i\over 4\pi \sqrt{3sm}}{s+m\over s-m}\mu_{s} e^{-i(s-m)\sigma_2}.
\end{eqnarray}
As expected form our configuration, the results for copies $(1)$ and $(3)$ are identical to the behaviour of a single two-twist operator, which was solved in general in \cite{Carson:2014xwa}. It is the behaviour of copy $(2)$ here which is new.

\subsection{Calculating the transition matrix}
While this twist configuration has not been solved before, the behaviour of copies $(1)$ and $(3)$ is the same as a single two-twist configuration since each copy only interacts with one of our twists. We can thus identify their transition matrices from \cite{Carson:2014xwa}.
\begin{eqnarray}
f^{(1)}_{ms} &=& \begin{cases}
\displaystyle{1\over \sqrt{3}}\delta_{ms} & s \in \mathbb{Z}\\
\displaystyle-{1\over 2\pi i\sqrt{3}}{1\over s-m}\mu_{-s} e^{-i(s-m)\sigma_1} & s \notin \mathbb{Z}
\end{cases} \nonumber \\
f^{(3)}_{ms} &=& \begin{cases}
\displaystyle{1\over \sqrt{3}}\delta_{ms} & s \in \mathbb{Z}\\
\displaystyle{1\over 2\pi i\sqrt{3}}{1\over s-m}\mu_{s} e^{-i(s-m)\sigma_2} & s \notin \mathbb{Z}.
\end{cases}
\end{eqnarray}

We now turn to $f^{(2)}$, which is where we must apply our method of combining transition behaviours for each individual two-twist operator. Noting that the intermediate state is on a doubly-wound CFT, we have:
\begin{eqnarray}
f^{(2)}_{ms} &=& \sum_{q \in \mathbb{Z}^+/2} f^{[(1),(2)]}_{mq}(\sigma_1)f^{[(1+2),(3)]}_{qs}(\sigma_2),
\end{eqnarray}
where here the copy indices in brackets are used to denote which copies are being twisted together. The expression $(1+2)$ refers to the intermediate doubly-wound CFT that results form the application of the twist at $\sigma_1$. All first-order transition matrices are known for two-twist operators\footnote{Of positive $SU(2)_L$ charge}, so we simply refer to \cite{Carson:2014xwa} for their continuum limit approximations.
\begin{eqnarray}
f^{[(1),(2)]}_{mq} &=& \begin{cases}
\displaystyle{1\over \sqrt{2}}\delta_{mq} & q \in \mathbb{Z}\\
\displaystyle{1\over 2\pi i\sqrt{2}}{1\over q-m}\mu_{q} e^{-i(q-m)\sigma_1} & q \notin \mathbb{Z}
\end{cases} \nonumber \\
f^{[(1+2),(3)]}_{qs} &=& \begin{cases}
\displaystyle\sqrt{2\over3}\delta_{qs} & s \in \mathbb{Z}\\
\displaystyle-{1\over 2\pi i\sqrt{6}}{1\over s-q}\mu_{s} e^{-i(s-q)\sigma_2} & s \notin \mathbb{Z}.
\end{cases}
\end{eqnarray}
We now break our computation into two cases.

\subsubsection*{Case 1: $s\in \mathbb{Z}$}
Here $f^{[(1+2),(3)]}$ is nonzero only for $q=s$, which means $q \in \mathbb{Z}$. Thus $f^{[(1)(2)]}$ is nonzero only for $m=q=s$. This leaves only one nonzero term in the sum.
\begin{eqnarray}
\left[f^{(2)}_{ms}\right]_{s\in \mathbb{Z}} &=& {1\over \sqrt{3}}\delta_{ms}.
\end{eqnarray}

\subsubsection*{Case 2: $s\notin \mathbb{Z}$}
Here we cannot have $s=q$ since the allowed values for the two indices only coincide at integers. We can however still have $q=m$. We separate out this term, which is the only nonzero term for integer $q$. This gives:
\begin{eqnarray}
\left[f^{(2)}_{qs}\right]_{s\notin\mathbb{Z}} &=& {\mu_s\over 4\pi \sqrt{3}}\left({i\over s-m} e^{-i(s-m)\sigma_2} - {1\over \pi}e^{im\sigma_1}e^{-is\sigma_2}\sum_{q\in \mathbb{Z} + {1\over2}, \, q>0} {e^{iq(\sigma_2-\sigma_1)}\over (s-q)(q-m)}\right).\qquad
\end{eqnarray}
As before, we will tackle this sum by using a large but finite cutoff.

\subsubsection*{A quick check}
Since there is no previous result to compare $f^{(2)}$ to, we instead multiply by $\alpha^{(2)}$ and use a large but finite cutoff to handle the sum. The product quickly reaches the identity thanks to the inverse square scaling of the summand. As with the previous configuration, a cutoff of $\Lambda = 2000$ is sufficient for $0.01\%$ accuracy.

\begin{figure}
 \begin{center}
  \includegraphics[width=0.8\textwidth]{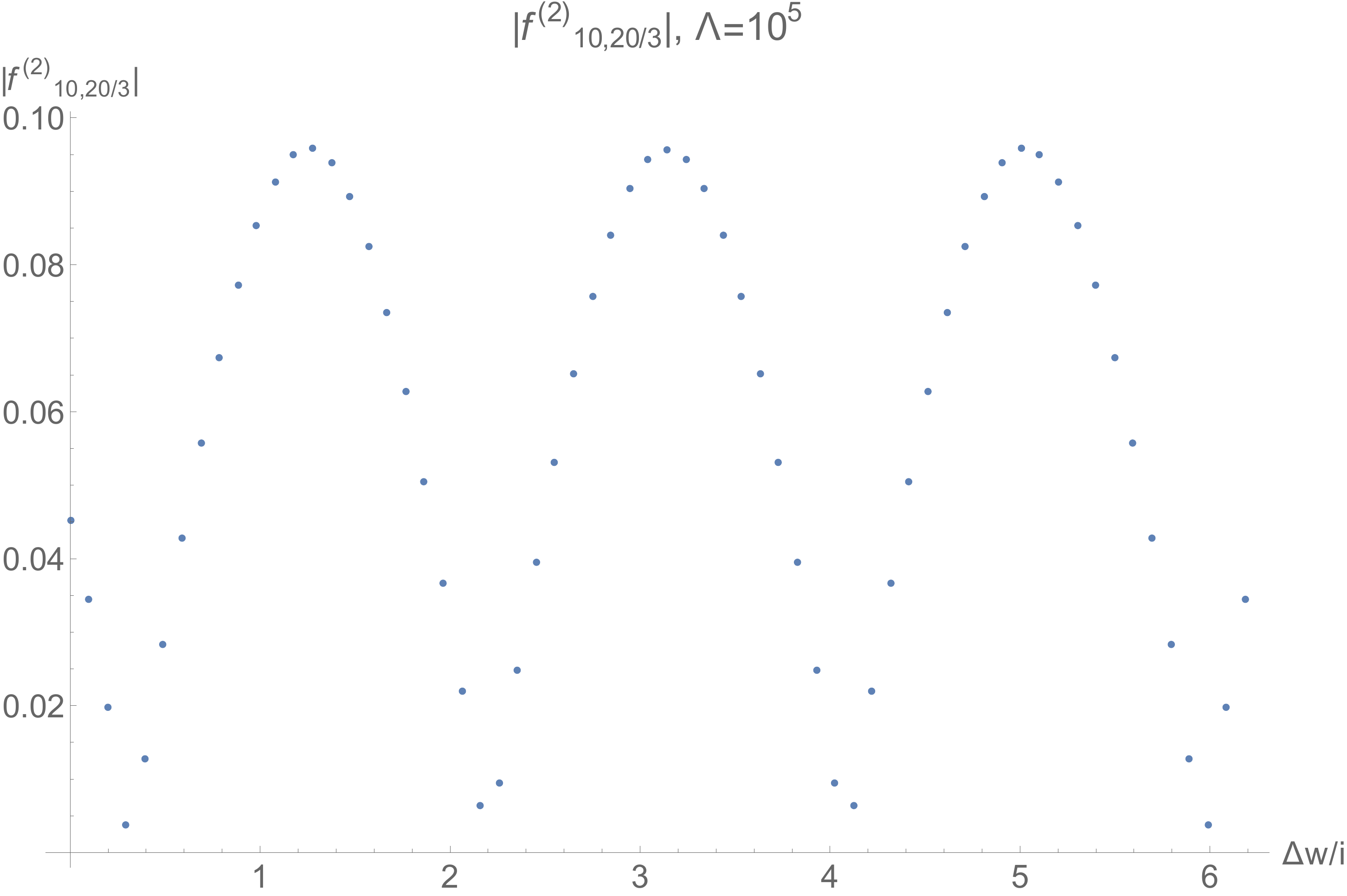}
 \end{center}
  \caption{Transition Amplitude as a function of separation for $f^{(2)}_{10,20/3}$.} \label{NewTransitionAmplitudeSeparationPlot}
\end{figure}

\begin{figure}
 \begin{center}
  \includegraphics[width=0.8\textwidth]{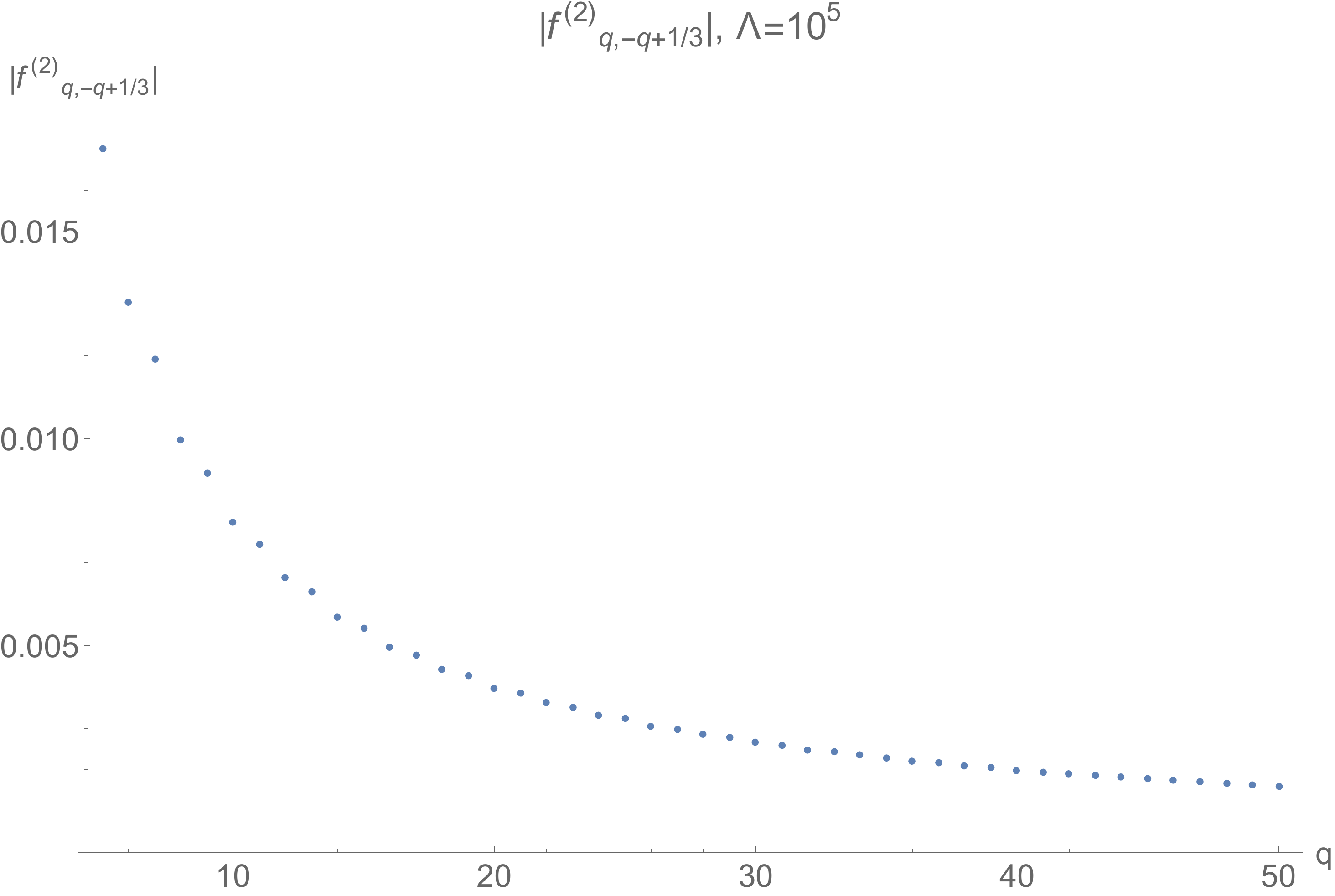}
 \end{center}
  \caption{Transition Amplitude as a function of its index for $f^{(2)}_{q,-q+1/3}$ at $\Delta w = \pi i$.} \label{TransitionAmplitudeFaloffPlot}
\end{figure}

The behaviour of $f^{(2)}_{ms}$ is shown in Figures \ref{NewTransitionAmplitudeSeparationPlot} and \ref{TransitionAmplitudeFaloffPlot}. As anticipated by \cite{Carson:2016cjj}, the continuum limit behaviour for $s \neq m$ scales as (for canonical modes):
\begin{eqnarray}
f^{(2)}_{ms} &\sim& {1\over (m-s)}\,h_f(m,s,\Delta w),\label{GeneralTransitionScaling}
\end{eqnarray}
where $h_f$ is a function oscillating in $\Delta w$ with amplitude independent of $m$ and $s$. This index falloff has been shown to hold for all cases of a single two-twist operator as well as for both second-order cases considered so far. We recommend this form of transition matrices as a first-guess fit for numeric work for any twist configurations in the continuum limit of the D1D5 CFT.

\subsection{The squeezed state matrix}
\begin{figure}
 \begin{center}
  \includegraphics[width=0.8\textwidth]{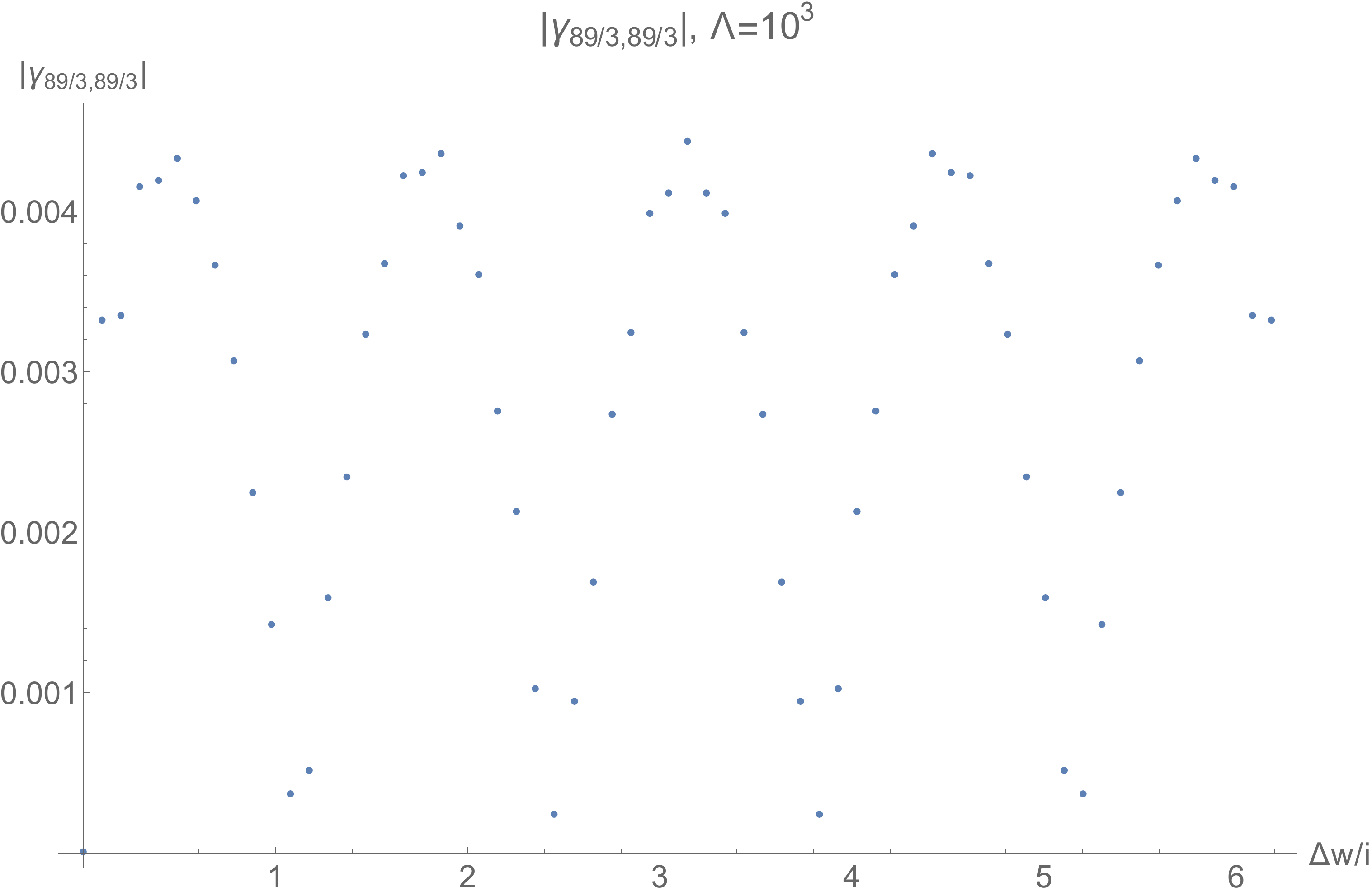}
 \end{center}
  \caption{Gamma as a function of separation for $\gamma_{89/3,89/3}$.} \label{NewGammaSeparationPlot}
\end{figure}

\begin{figure}
 \begin{center}
  \includegraphics[width=0.8\textwidth]{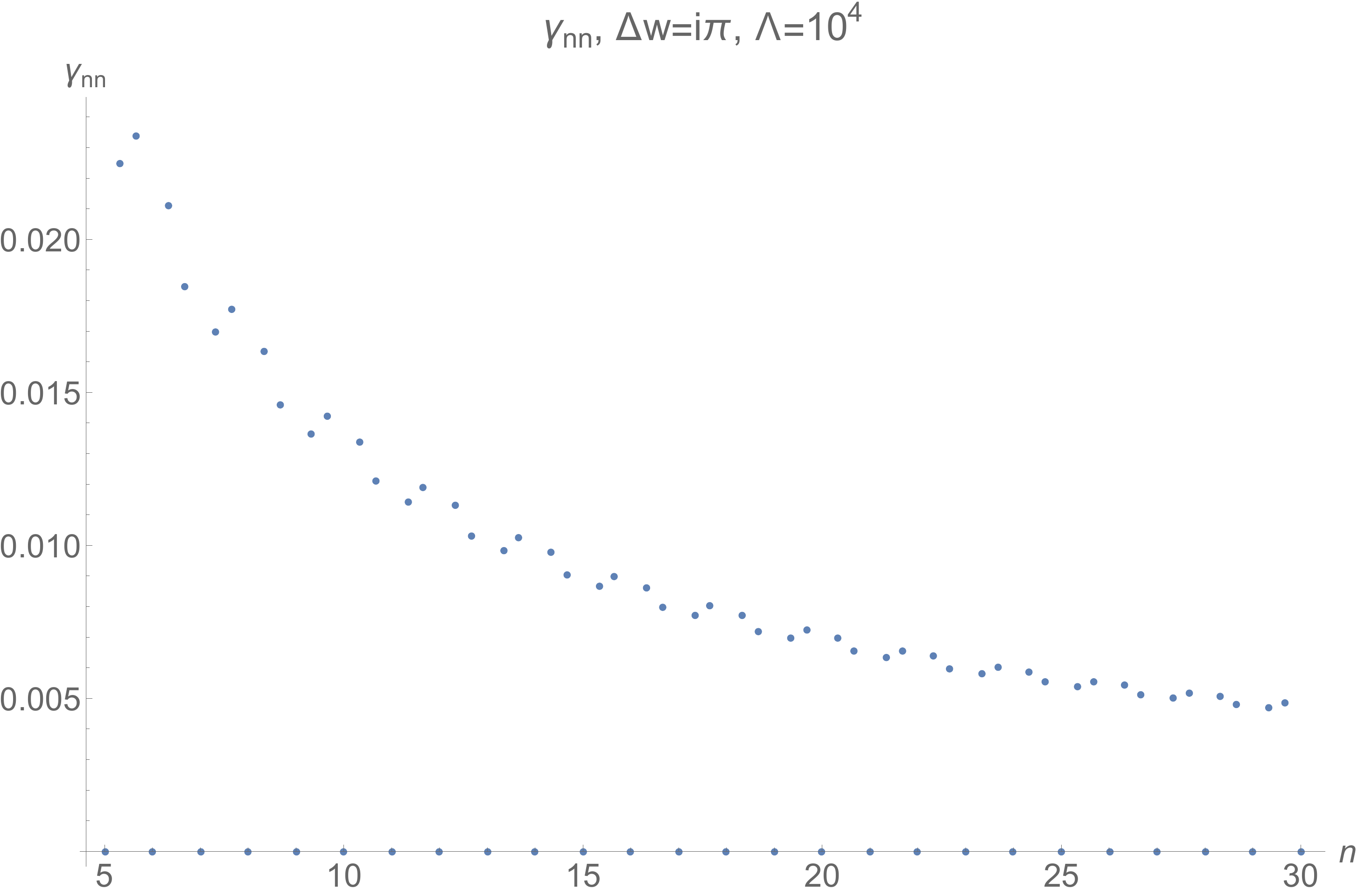}
 \end{center}
  \caption{Gamma as a function of its index for $\gamma_{s,s}$ at $\Delta w = \pi i$.} \label{NewGammaFalloffPlot}
\end{figure}

We can now combine our transition matrix with $\beta$ to obtain the matrix $\gamma$ that characterizes the squeezed state. All three initial copies contribute.
\begin{eqnarray}
\gamma_{ss'} &=& \left[f^T\beta\right]_{ss'} ~=~ \sum_{m>0}\sum_{(i)}f^{(i)}_{ms}\beta^{(i)}_{ms'}.
\end{eqnarray}
There are technically two sums at work here, as a sum is required for the intermediate calculation of $f^{(2)}$. As usual we terminate each sum at some large cutoff. The resulting $\gamma_{ss'}$ behaviour is shown in Figures \ref{NewGammaSeparationPlot} and \ref{NewGammaFalloffPlot}. Once again, the anticipated falloff from \cite{Carson:2016cjj} is a good approximation for the continuum limit.
\begin{eqnarray}
\gamma_{ss'} &\sim& {1\over s+s'} h_{\gamma}(s,s',\Delta w),\label{GeneralGammaScaling}
\end{eqnarray}
where $h_{\gamma}$ is a different function oscillating in $\Delta w$ with amplitude independent of $s$ and $s'$. We therefore recommend this falloff form for the squeezed state coefficient as a first-guess fit for numeric work with any twist configurations in the continuum limit of the D1D5 CFT.

\section{Discussion}\label{SectionDiscussion}

Our goal here was to further investigate the effect of twist operators on quantum states in the D1D5 orbifold CFT, in order to bring us closer to the general goal of understanding how thermalization works in the microscopic picture of black holes. We wished to find results in the continuum limit that went beyond the already known twist 2 results found previously. The method we presented here, using component twists, has the advantage of being applicable to higher twist configurations, unlike the Lunin-Mathur method.

We used knowledge of Bogoliubov transformations and the fact that twist two operators $\hat{\sigma}_2$ can be used to build up higher-twist operators $\hat{\sigma}_n$ to find the transition matrices $f$ and squeezed state coefficient matrices $\gamma$ associated to states created by $\hat{\sigma}_n$. To show our method works, we checked our component twist method by comparing to a previous result calculated with Lunin-Mathur technology.

With our method verified, we explicitly calculated a new configuration and found that the scaling predicted in \cite{Carson:2016cjj} holds true: we see good agreement with equations (\ref{GeneralTransitionScaling}) for the transition matrices $f$ and (\ref{GeneralGammaScaling}) for the squeezed state coefficients $\gamma$. We expect that the form of these matrices will hold in the continuum limit for any general twist operator configurations. This conclusion is based on our component twist method, the forms of the twist 2 transition matrices, and simple power counting.

We have seen, using calculations in the D1D5 orbifold CFT, that it is possible to build up the $f$ matrices from components and to obtain the $\gamma$ matrices from them. The only part that made our calculations specific to this CFT was the form of the twist two transition amplitudes. Twist operators in other orbifold CFTs might have a similar decomposition. 

Twist operators also show up in other orbifold CFTs that are not themselves holographic but are useful for computing quantities of physical interest in holographic CFTs. In particular, cyclic orbifolds show up when using the replica trick for computing entanglement entropy via Renyi entropies in quantum field theories \cite{Calabrese:2004eu,Giusto:2014aba,Lashkari:2014yva,Lin:2016fqk,Balasubramanian:2016xho}. The entanglement entropy of multiple regions is written as a correlation function of multiple pairs of twist operators. Better understanding the effect of applying twist operators in the prototype D1D5 orbifold CFT may assist with learning about information theoretic quantities of interest for quantum gravity. 

\section*{Acknowledgements}

We would like to thank Samir D. Mathur for useful discussions. This work is supported by a Discovery Grant from the Natural Sciences and Engineering Research Council of Canada. Computations were performed on the GPC supercomputer at the SciNet HPC Consortium. SciNet is funded by: the Canada Foundation for Innovation under the auspices of Compute Canada; the Government of Ontario; Ontario Research Fund - Research Excellence; and the University of Toronto.

\end{document}